# Competing multiferroic phases in monolayer and few-layer NiI$_2$

Nanshu Liu[1,2], Cong Wang[1,2], Changlin Yan[1,2], Changsong Xu[3], Jun Hu[4*], Yanning Zhang[5*] and Wei Ji[1,2,5*]

[1]*Beijing Key Laboratory of Optoelectronic Functional Materials & Micro-Nano Devices, Department of Physics, Renmin University of China, Beijing 100872, China*
[2]*Key Laboratory of Quantum State Construction and Manipulation (Ministry of Education), Renmin University of China, Beijing, 100872, China*
[3]*Key Laboratory of Computational Physical Sciences (Ministry of Education), Institute of Computational Physical Sciences, State Key Laboratory of Surface Physics and Department of Physics, Fudan University, Shanghai 200433, China*
[4]*School of Physical Science and Technology, Ningbo University, Ningbo, 315211, China*
[5]*Institute of Fundamental and Frontier Sciences, University of Electronic Science and Technology of China, Chengdu, 610054, China*
*Corresponding authors. E-mail: hujun2@nbu.edu.cn; yanningz@uestc.edu.cn; wji@ruc.edu.cn.

**Abstract:** A recent experiment reported type-II multiferroicity in monolayer (ML) NiI$_2$ based on a presumed spiral magnetic configuration (Spiral-B), which is, as we found here, under debate in the ML limit. Freestanding ML NiI$_2$ breaks its $C_3$ symmetry, as it prefers a striped antiferromagnetic order (AABB-AFM) along with an intralayer antiferroelectric (AFE) order. However, substrate confinement may preserve the $C_3$ symmetry and/or apply tensile strain to the ML. This leads to another spiral magnetic order (Spiral-IV$^X$), while 2L shows a different order (Spiral-V$^X$) and Spiral-B dominates in thicker layers. Thus, three multiferroic phases, namely, Spiral-B+FE, Spiral-IV$^X$+FE, Spiral-V$^X$+FE, and an anti-multiferroic AABB-AFM+AFE one, show layer-thickness-dependent and geometry-dependent dominance, ascribed to competitions among thickness-dependent Kitaev, biquadratic, and Heisenberg spin–exchange interactions and single-ion magnetic anisotropy. Our theoretical results clarify the debate on the multiferroicity of ML NiI$_2$ and shed light on the role of layer-stacking-induced changes in noncollinear spin–exchange interactions and magnetic anisotropy in thickness-dependent magnetism.

Magnetoelectric (ME) effects enable the manipulation of magnetic (electric) properties using electric (magnetic) fields, which is of interest in terms of both fundamental physics and potential spintronic applications [1, 2]. ME manipulations can be achieved by multiferroic materials exhibiting magnetic and electric orders [3]. Usually, the electric polarization of a type-II multiferroic material is induced by a magnetic order spontaneously through the inverse Dzyaloshinskii–Moriya (D–M) interaction [4-7]. However, the fact that the identification of type-II multiferroics in atomically thin van der Waals (vdW) monolayers (MLs) is still under debate. $NiI_2$ is a highly promising candidate for ML multiferroicity. Its bulk form undergoes two successive magnetic phase transitions [8, 9] from a paramagnetic (PM) phase to an interlayer antiferromagnetic (AFM) phase at $T_{N1}$ = 76 K and then to a spiral magnetic phase below $T_{N2}$ = 59.5 K [8]. The AFM-to-spiral transition is accompanied by breaks of both rotational and inversion symmetries, which results in electric polarization through the inverse D–M interaction. This effect is reflected in second harmonic generation (SHG) [10] and birefringence signals [11].

Monolayer $NiI_2$ on an hBN substrate was recently shown to exhibit multiferroicity under 20 K as its enhanced SHG signal strength [11], ascribed to a broken inversion symmetry of either magnetic or geometric (electrical) origin. However, the magnetic configuration of ML $NiI_2$ below 20 K remains an open issue, as no direct experimental measurement has been successfully conducted and previous density functional theory (DFT) predictions are under debate among ferromagnetic (FM) [12-14], helimagnetic [15, 16], and AFM [17] configurations. Although not conclusive, they are indeed different from the bulk groundstate. If the SHG enhancement originates from the magnetic contribution, it cannot guarantee the formation of additional electric polarization below 20 K [11]. This concern about the claimed multiferroicity was reinforced by the fact that no SHG enhancement was observable in ML $NiI_2$ on a $SiO_2$ substrate [10]. Therefore, the assertion of multiferroic ML $NiI_2$ requires further verification beyond the SHG measurement [11], which raises the question of whether type-II multiferroicity persists in the ML limit.

In this work, we examine the evolution of the magnetic groundstate and electric polarization of $NiI_2$ from the bulk to the ML using DFT calculations. We suggest a tentative layer-dependent magnetic phase diagram that illustrates the competition among, at least, four magnetic phases, with three spiral ones (one collinear) being (anti-)ferroelectric and induced by the inverse D–M interaction. The magnetic groundstate of ML highly depends on the in-plane geometry to stabilize the competing Spiral-$IV^X$ ($q_{IV}^X$) and AABB-AFM configurations under different strains. We additionally demonstrate that these phase changes are driven by the competition among layer-thickness and local-geometry-dependent (non-)collinear Kitaev, biquadratic, and isotropic Heisenberg spin–exchange interactions and single-ion anisotropy. We also construct $NiI_2$/hBN and $NiI_2$/$SiO_2$ heterostructures to consistently explain the SHG signal of ML $NiI_2$ over 20 K obtained by two pioneering experiments [10, 11].

The spin–exchange coupling parameters were extracted based on the following Hamiltonian[18]

$$H = -\frac{1}{2}\left[\sum_{i,j;x,y,z} J_{ij} \boldsymbol{S}_i \cdot \boldsymbol{S}_j + \sum_{i,j;x,y,z} J_{ij}^{\perp} \boldsymbol{S}_i \cdot \boldsymbol{S}_j + \sum_{i,j;\alpha,\beta(\gamma)} \left(\lambda_\alpha S_i^\alpha S_j^\alpha + \lambda_\beta S_i^\beta S_j^\beta + \lambda_\gamma S_i^\gamma S_j^\gamma\right) + \sum_{i,j} B(\boldsymbol{S}_i \cdot \boldsymbol{S}_j)^2 + 2\sum_{i;x,y,z} A_z S_i^2\right] = -\frac{1}{2}[\sum_{i,j} J_{ij}\boldsymbol{S}_i \cdot \boldsymbol{S}_j + \sum_{i,j} J_{ij}^{\perp}\boldsymbol{S}_i \cdot \boldsymbol{S}_j + \sum_{i,j}(K_{ij}^\gamma S_i^\gamma S_j^\gamma) + \sum_{i,j} B(\boldsymbol{S}_i \cdot \boldsymbol{S}_j)^2 + 2\sum_i A_z S_z^2],$$

where $A_z$ represents the single-ion anisotropy, $J_{ij}$ and $J_{ij}^{\perp}$ are the intra- and inter-layer isotropic Heisenberg exchange parameters, $B$ and $K_{ij}^\gamma$ are the biquadratic and collinear Kitaev interaction parameters [19]. We followed the procedures used in our previous calculations [20, 21] and have included the details in the Appendix-A.

Bulk $NiI_2$ crystal has a rhombohedral structure in space group R$\bar{3}$m at room temperature (Fig. 1a), comprised of triangularly arranged $Ni^{2+}$ cations ($3d^8$, $S=1$) and coordinating I anions. We used a $1 \times \sqrt{3} \times 1$ supercell to more clearly show magnetic configurations in Fig. 1e. The calculated lattice constants $a = 3.926$ Å, $b = 6.790$ Å, and $c$ = 19.744 Å of bulk NiI$_2$ crystal in the experimentally observed spiral order (Spiral-B), consistent with the experimental values of $a$ = 3.919 Å, $b$ = 6.765 Å, and $c$ = 19.635 Å [22]. Spiral-B exhibits a propagation vector $q_B$ = (0, 0.138, 1.457) (Figs. 1c and 1d) below $T_{N2}$ = 59.5 K [11, 22]. Note that in the coordinates defined in the literature [11, 22], $q_B$ = (0.138, 0, 1.457). Our DFT calculations reproduced this magnetic order suggested in experiments [11, 23] and obtained by a spin Hamiltonian [24] in which Spiral-B exhibits the lowest energy compared to the five collinear magnetic configurations (Fig. 6 in Appendix-B) and 55 other spiral configurations with different $q$ values (Figs. 1e and 1f). The Spiral-B groundstate is robust regardless of the preservation of the $C_3$ symmetry (Table I), consideration of spin-orbit coupling (SOC) (Fig. 7), and choice of on-site Coulomb interaction ($U_{eff}$) values (Fig. 8). We are thus confident of the reliability of our results for ML or few-layer NiI$_2$.

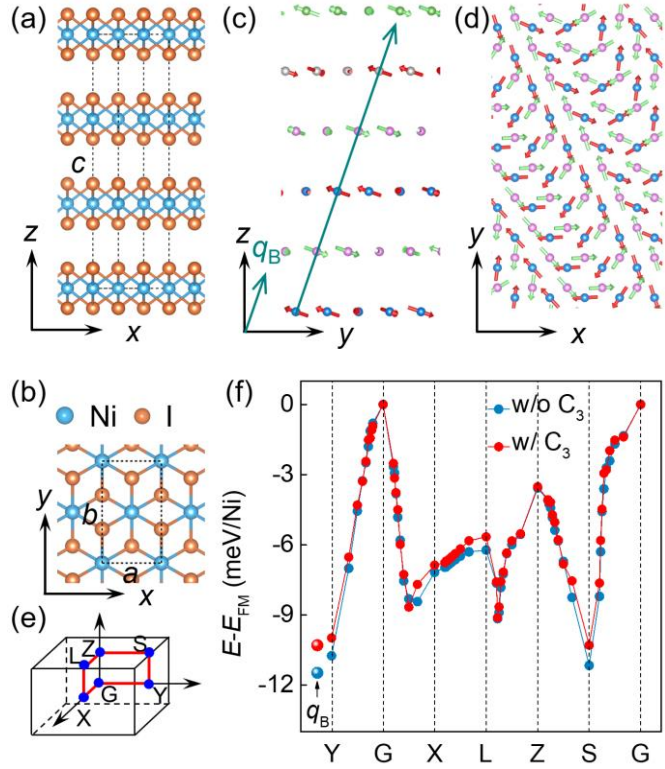

FIG. 1. Side (a) and top (b) views of the atomic structure of bulk NiI$_2$. Schematics of Spiral-B in side (c) and top (d) views. Only two Ni layers were plotted in (d) to more clearly show the magnetic moments. (e) $q$-path considered in the first Brillouin zone of a $1 \times \sqrt{3} \times 1$ supercell. (f) Relative energies of all considered spiral orders with (red) and without (blue) preserved $C_3$ structural symmetry.

The magnetism of ML and few-layer NiI$_2$ is more complicated than bulk NiI$_2$

and difficult to be captured by a spin Hamiltonian. We considered 25 collinear and 71 non-collinear magnetic orders using the supercell model and a $10 \times 10$ $q$-mesh for the generalized Bloch theorem (gBT) model to explore the layer-dependent magnetic groundstates. We focus on the geometries showing (ML-C3) and not showing (ML-NC3) the $C_3$ symmetry for the ML because the geometric symmetry is more easily modulated by the substrate [25-27]. For thicker layers, up to 4L, we concentrate on constraint-free geometries as the substrate constraint rapidly relaxes in thicker layers [28].

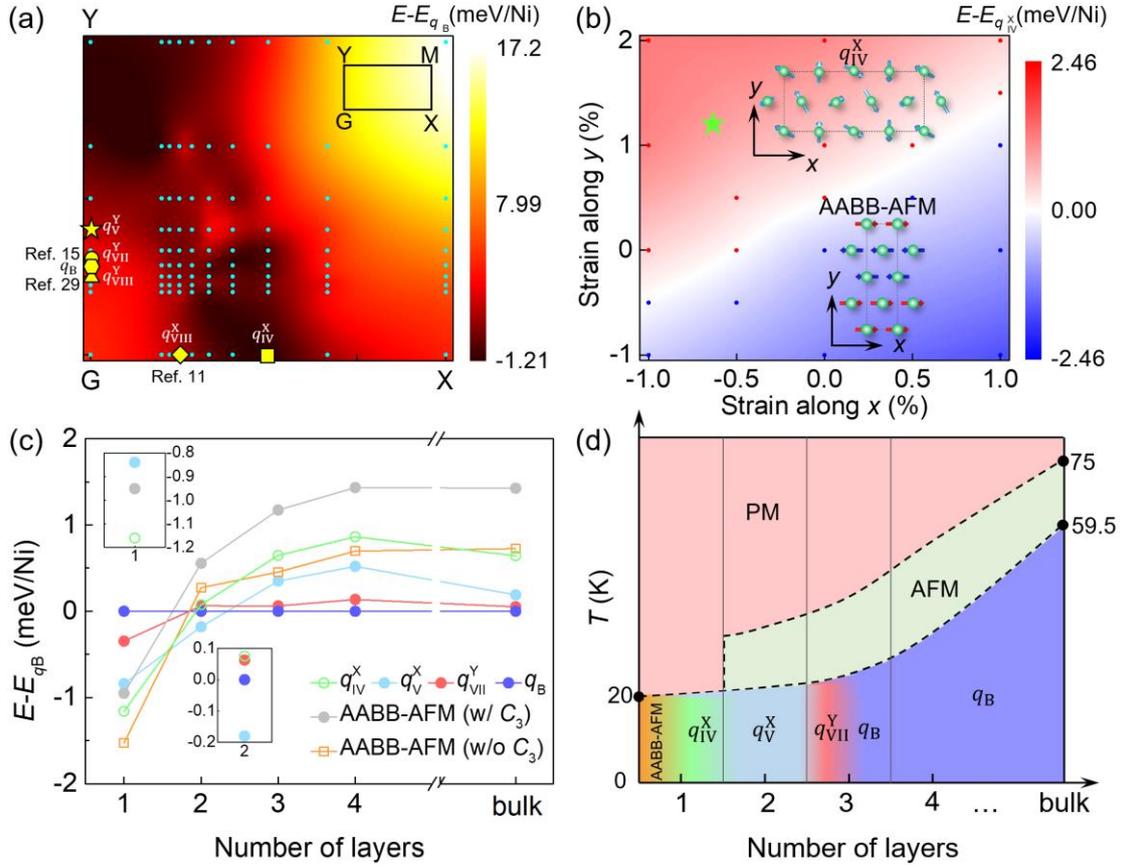

FIG. 2. (a) Energy difference mapping between spin-spiral and Spiral-B orders for ML-C3. Inset: $q$-path in the first Brillouin zone of a $1 \times \sqrt{3}$ supercell. The spiral vectors in bulk NiI$_2$ ($q_B$), Spiral-IV$^X$ ($q_{IV}^X$), Spiral-V$^Y$ ($q_V^Y$), Spiral-VIII$^X$ ($q_{VIII}^X$) from Ref. [11], Spiral-VII$^Y$ ($q_{VII}^Y$) from Ref. [15] and Spiral-VIII$^Y$ ($q_{VIII}^Y$) from Ref. [29] are indicated by yellow polygons. (b) Energy difference between AABB-AFM and Spiral-IV$^X$ versus epitaxial strains along the $x$ and $y$ directions for ML-NC3. Insets: top views of Spiral-IV$^X$ and AABB-AFM orders. The green star denotes the strain values of ML NiI$_2$ being applied from an $h$BN substrate. The original data was represented by colored dots in (a) and (b). (c) Layer-dependent energy difference between the spin–spiral and Spiral-B orders for both structures. Insets: zoomed-in energies for ML and 2L NiI$_2$. (d) Schematic magnetic phase diagrams for different NiI$_2$ layers

versus temperature, with PM and AFM representing paramagnetic and interlayer AFM states, respectively. The transition temperatures were taken from an experimental work [30].

Figure 2a (Figure 9 in Appendix-C) plots an energy map (profile) for the $q$-mesh (differently sized supercells) of ML-C3 NiI$_2$, where the $C_3$ symmetry is preserved under constraint. In the following, all energy comparisons were based on the results by constructing supercells including SOC. Both plots indicate a new spiral configuration (Spiral-IV$^X$, upper inset in Fig. 2b), the magnetic moments of which align in the Ni-I plane and follow the propagating vector $q_{IV}^X$ = (0.250, 0, 0) (across each 4 × $\sqrt{3}$ supercell, Figs. 10a and 10b), consistent with a recently spin-polarized scanning tunneling microscopy (SP-STM) observation [31], and same as a previous work but with magnetic moments rotating in the $yz$ plane [32]. Spiral-IV$^X$ shares the same propagation direction with the spiral order discussed in Ref. [11] (Spiral-VIII$^X$) where the propagating period twice that of Spiral-IV$^X$, namely eight-unit cells (Fig. 11). Spiral-IV$^X$ is, at least, 0.16 meV/Ni more stable than the other spiral orders listed in Table II (e.g., Spiral-V$^Y$, Spiral-VII$^Y$, and Spiral-B; see Fig. 2c) and a collinear AABB-AFM order (lower inset in Fig. 2b), while other configurations (Figs. 12 and 13) are even less stable than the abovementioned five. Spiral-IV$^X$ is also more stable by at least 0.81 to 1.75 meV/Ni than the two recently theoretically suggested [15, 29] and one experimentally observed [33] spiral orders.

In the constrain-free case, the AABB-AFM structure breaks the $C_3$ symmetry (ML-NC3), which is more stable than Spiral-IV$^X$ by 0.35 meV/Ni (Table II and Fig. 2c). This suggests that external strain plays a role in tuning their relative stability. As shown in a phase diagram in Fig. 2b, Spiral-IV$^X$ is substantially stabilized under in-plane compressive strain along the $x$ direction and/or tensile strain along the $y$ direction (red zone in Fig. 2b). Order Spiral-IV$^X$ is, at least, 0.12 meV/Ni more stable than Spiral-V$^Y$ and Spiral-VII$^Y$ in the whole considered strain range, while order AABB-AFM becomes even more stable than order Spiral-IV$^X$ in certain strain regions (Fig. 14 in Appendix-D). Thus, the magnetic groundstate was compared between Spiral-IV$^X$ and AABB-AFM orders for simplicity.

For constrain-free 2L NiI$_2$, the AABB-AFM order becomes less stable compared to those spiral orders. A new spiral order, Spiral-V$^X$ ($q_V^X$), emerges and is 0.26 meV/Ni more stable than Spiral-IV$^X$. Spiral-V$^X$ also propagates in the *x* direction, following vector $q_V^X$ = (0.20, 0, 0), almost degenerated with the order observed in the SP-STM experiment [31], where a spiral order exhibits 5.01*a*, deviating by 7° from the *x* direction. Spiral-V$^X$ is more stable than Spiral-VII$^Y$ ($q_{VII}^Y$) and Spiral-B by 0.24 and 0.18 meV/Ni, respectively (Table III in Appendix-E, Fig. 2c), nearly energetically undistinguishable. Here, Spiral-VII$^Y$ propagates along the *y* direction across a 1 × 7 supercell and represents the in-plane projection of Spiral-B, indicating that the interlayer non-collinear spin–spin interactions are rather weak compared to their in-plane counterparts. Breakdown or preservation of the $C_3$ symmetry does not essentially change the relative stability of these spiral orders (Table III).

Spiral-VII$^Y$ and Spiral-B are still energetically undistinguishable in 3L, but over 0.35 meV/Ni more stable than Spiral-IV$^X$ and Spiral-V$^X$ (Table IV). In 4L and thicker layers, the interlayer spin–spin interactions play a more crucial role as Spiral-B becomes the groundstate by at least 0.14 meV/Ni (Table V). These results depict a layer-dependent competition of one collinear and four spiral orders within four layers, as schematically summarized in the magnetic phase diagram (Fig. 2d).

We examined the impact of $U_{\text{eff}}$ and functional on relative energies of those competing configurations. The $U_{\text{eff}}$ values up to 5.4 eV don't affect their relative stability (Fig. 15a in Appendix-F) and are already larger than the values used in the literature [34] and obtained by a linear response method [35]. Their stability was also verified using the Perdew–Burke–Ernzerhof (PBE), PBE-D3, Revised PBE, and HSE06 hybrid functionals considering SOC (Fig. 15b), despite numerical discrepancies between the results obtained from HSE06 and other functionals.

Orders Spiral-IV$^X$, Spiral-V$^X$, and Spiral-B (Spiral-VII$^Y$) induce electric polarization through the inverse D–M interaction, represented by $\boldsymbol{P} \parallel \hat{\boldsymbol{e}} \times \boldsymbol{q}$, where $\hat{\boldsymbol{e}}$ denotes the rotational axis of the spiral spins [36]. In Fig. 3a, the clockwise-rotating spins propagating along the *x* direction in the $q_{IV}^X$ generate an in-plane electric polarization in the $y$ direction ($P_y$), perpendicular to bulk in the experiment [8]. A switchable polarization vector characterizes an FE material rather than an electret. Figure 3b illustrates a likely intermediate configuration in a switching process of rotating spins propagating from the $+x$ ($q_{IV}^X$) to the $-x$ direction ($-q_{IV}^X$, Fig. 3c). The change in the propagating direction switches the electric polarization from the $+y$ (Fig. 3a) to the $-y$ (Fig. 3c), surmounting an energy barrier of ~6 meV in ML-C3 NiI$_2$ (Fig. 16a in Appendix-G), which is comparable to that of FE ML SnSe (3.76 meV) [37] but smaller than FE ML Bi (43 meV) [38]. The electric polarizations for Spiral-VII$^Y$ and Spiral-B are in the $x$ direction ($P_x$) for their propagating directions along the $y$ direction. In Fig. 3d, the bulk value of 0.90 pC/m (for Spiral-B) gradually drops to 0.24 pC/m in ML-C3 (Spiral-IV$^X$), comparable to the values for ML FeOCl (~ 0.39 pC/m) [39] and ML Hf$_2$VC$_2$F$_2$ (0.29 pC/m) [40].

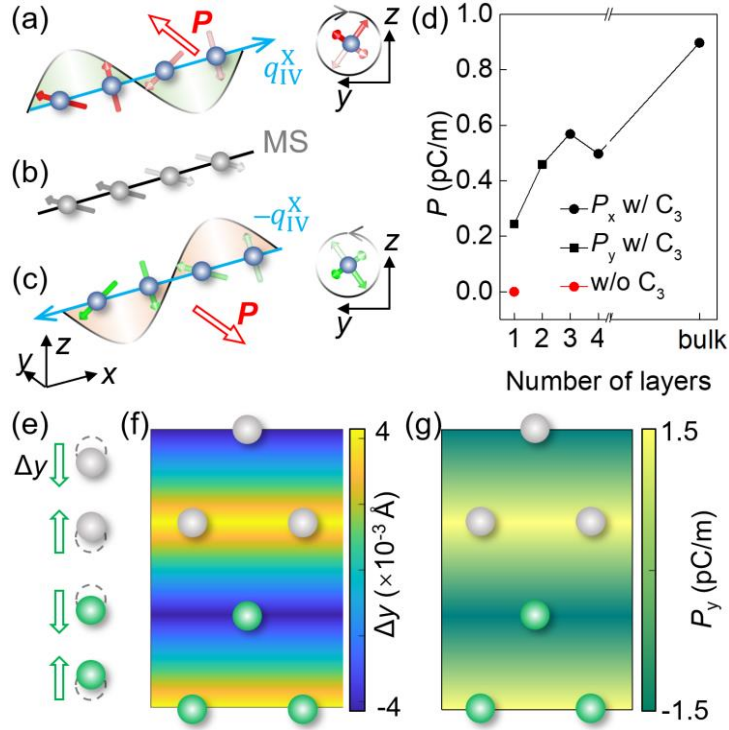

FIG. 3. Schematic plots for a spin canting process from (a) a clockwise propagation in Spiral-IV$^X$ ($q_{IV}^X$) to (c) a counterclockwise one ($-q_{IV}^X$) through (b) a likely metastable state (MS). The (counter-)clockwise spin spiral generates an in-plane electric polarization (anti-)parallel to the $y$ direction. (d) Theoretical electric polarization versus the number of NiI$_2$ layers. (e–f) Displacement of Ni atoms in AABB-AFM along the $y$ direction in ML-NC3. (g) Electric polarization induced by displacements of Ni atoms in the AABB-AFM state.

However, the AABB-AFM order in ML-NC3 eliminates the total electric polarization and exhibits intralayer antiferroelectricity for the shrank lattice along the $y$-axis accompanying with Ni atoms relaxation (Table VI and Fig. 3e). The Ni atoms in the same AFM stripe move oppositely by 0.004 Å in the $\pm y$ directions (Fig. 3f), yielding a dipole moment of $\pm 1.5$ pC/m (Fig. 3g). The switching barrier (Fig. 16b) between the two AFE configurations is ~6 meV, comparable to the FE barrier of ML Spiral-IV$^X$.

In short, few-layer NiI$_2$ exhibits at least four competing multiferroic phases: the newly found Spiral-IV$^X$ + FE, Spiral-V$^X$ + FE, and AABB-AFM + AFE states and the previously known Spiral-B + FE state. Thus, ML NiI$_2$ is indeed a type-II multiferroic material, but with tunable multiferroicity between AFM + AFE and Spiral + FE by the in-plane geometry. This helps explain the seemingly contradictory experimental results [10, 11], where ML NiI$_2$ was supported on SiO$_2$ [10] and $h$BN [11] substrates, suggesting the substrate may play a role in affecting their in-plane geometry.

The ML NiI$_2$ was epitaxially grown on a $h$-BN substrate [11], while the structural details of ML-NiI$_2$ on the $h$-BN substrate remain unknown in experiments. We thus theoretically considered the ML-NiI$_2$/$h$-BN interface exhibiting the smallest lattice mismatch and lowest total energy [41]. We used an ML NiI$_2$ 10×4$\sqrt{3}$ / ML $h$BN 9$\sqrt{3}$×11-R30 rectangular supercell, in which the $h$BN substrate applies a compressive strain of −0.6 % along the $x$ direction and a tensile strain of ~1.3 % along the $y$ to ML NiI$_2$. This supercell is 1.43 and 2.48 meV/Ni more stable than the two configurations (Fig. 17 in Appendix-H) exhibiting the second and third smallest interfacial strains, which are in the tensile strain region. The Ni layer (Fig. 18a) and the two I sublayers (Fig. 19) exhibit out-of-plane corrugations varying up to 0.06 Å. Nonuniform in-plane strains further break the inversion symmetry in the $x$ (Fig. 18c) and $y$ (Fig. 18d) directions. Moreover, explicit interfacial charge transfer from the BN layer to the interfacial I layer leads to an out-of-plane electric polarization (Figs. 18e and 18f). Therefore, ML NiI$_2$ on $h$BN simultaneously breaks its structural inversion symmetry, corresponding to the observable SHG signals above 20 K in Ref. [11].

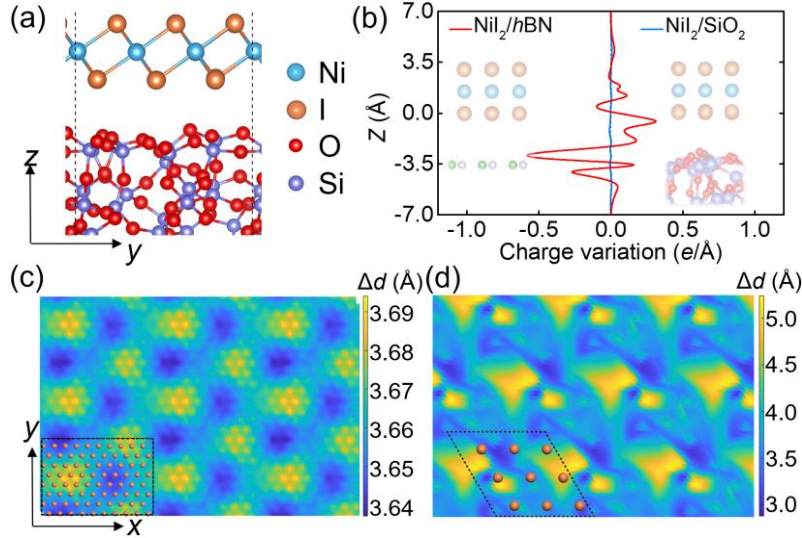

FIG. 4. (a) Side view of the optimized geometrical structure of a NiI$_2$ ML deposited on an amorphous SiO$_2$ substrate. (b) Line profiles along the $z$ direction of charge variations for the NiI$_2$/$h$-BN and NiI$_2$/SiO$_2$ heterostructures. The $z$-coordinates of the interfacial I atoms were set to zero. 2D mappings of spatial variation of vertical distances between lower I atoms and (c) top-layer B or N atoms of the $h$-BN substrate and (d) surface O atoms of the SiO$_2$ substrate.

These strong modifications from the $h$BN to ML NiI$_2$ indicate their strong interactions. NiI$_2$ is, most likely, prone to maintain its C$_3$ symmetry on $h$BN due to the confinement from the C$_3$ symmetrized $h$BN. For ML-C3, Spiral-IV$^X$ is preferred over AABB-AFM and further stabilized under biaxial compressive strains and electron doping (Figs. 14b and 20 in Appendix-I). Thus, the SHG signal could be further enhanced by the additional in-plane electric polarization induced by the noncollinear Spiral-IV$^X$ order formed below 20 K [11]. However, the amorphous SiO$_2$ substrate is a well-saturated substrate that exhibits weak interactions with its supporting MLs [42]. The validity of this statement for ML NiI$_2$ was supported by our theoretical comparison of the NiI$_2$/SiO$_2$ and NiI$_2$/$h$BN interfaces (Fig. 4 and Appendix-J), in which the NiI$_2$/SiO$_2$ interface shows negligible interfacial charge variations (Fig. 4b) and larger interfacial distance variations (Figs. 4c and 4d). The SiO$_2$ substrate thus interacts more weakly to NiI$_2$, leading the NiI$_2$ overlayer more close to its freestanding form. Therefore, ML NiI$_2$ placed on amorphous SiO$_2$, most likely, favors the collinear AABB-AFM order and thus shows no temperature-dependent SHG enhancement [10].

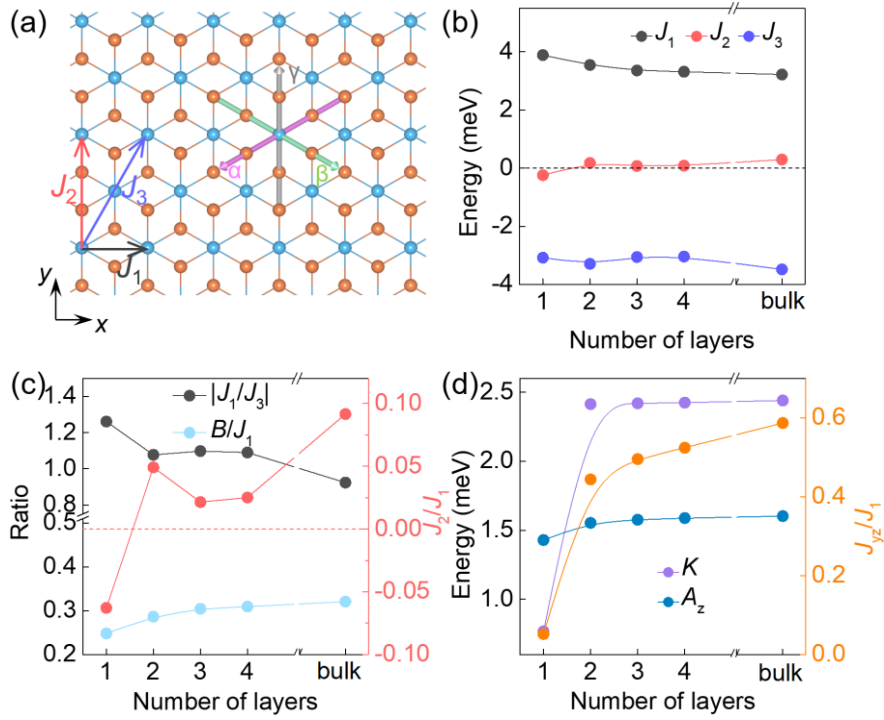

FIG. 5. (a) Schematic illustration of the Kitaev basis {$\alpha\beta\gamma$} (magenta, green, and grey arrows) and the intralayer isotropic first to third nearest-neighbor Heisenberg spin-exchange parameters $J_1$ to $J_3$. (b) Layer-dependent evolution of $J_1$, $J_2$, and $J_3$. (c) Ratios of $|J_1/J_3|$, $J_2/J_1$, and $B/J_1$. (d) Non-collinear $J_{yz}$ over $J_1$ ($J_{yz}/J_1$), Kitaev $K$ and single-ion anisotropy $A_z$.

The remaining issue is why ML and few-layer NiI$_2$ hosting at least four competing states. We plotted the layer dependence of various magnetic interactions (Fig. 5). Parallel coupled Heisenberg $J_1$ (black in Fig. 5a) nearly maintains its bulk value of 3.22 meV down to 3L and then rapidly increases to 3.55 meV in 2L and 3.89 meV in ML (Fig. 5b). Anti-parallel coupled $J_3$ (blue in Fig. 5a) exhibits almost the opposite trend (Fig. 5b). Their competition could lead to non-collinear spin spiral states if $|J_1/J_3| < 4$ [43]. The DFT revealed that $|J_1/J_3|$ ranges from 0.92~1.26 for NiI$_2$ layers (Fig. 5c) and fits this criterion well, while the strongest (weakest) frustration occurs in bulk (ML) where the ratio reaches its minimum (maximum) among all considered layer thicknesses.

Two-site anisotropy $J_{yz}$ characterizes the preference of the direction of magnetic moments canting from the $xy$-plane to the $z$-axis [29]. The bulk NiI$_2$ exhibits $J_{yz}$ = 1.90 meV and the largest $J_{yz}/J_1$ ratio of 0.59 (Fig. 5d), indicating the preferred out-of-plane (OOP) orientations of the magnetic moments in thicker layers, consistent to their OOP easy axes (Figs. 21b and 21c in Appendix-K) and in-plane one in ML (Fig. 21e). Single-ion anisotropy $A_z$, favoring the moments oriented in the *z*-axis, follows the same trend and reaches its maximum positive value of 1.60 meV (Fig. 5d) in bulk NiI$_2$. A considerably large Kitaev interaction $K$ = 2.44 meV (see Fig. 5d) confines magnetic moments in the *α–β* plane (Fig. 1c), which leads to Spiral-B being more stable than its *xy*-plane projection Spiral-VII$^Y$. The moments in Spiral-B are slightly off the *α–β* plane owing to competition between $K$ and $A_z$.

For layers thinning from bulk down to 4L–2L, the ratio between the competing FM $J_2$ and $J_1$ ($J_2/J_1$) decreases from approximately 0.03 (4L) to 0.02 (3L) and then increases to 0.05 (2L) (Fig. 5c), which results in the emergence of a spatially smaller spiral order (Spiral-V$^X$) in 2L. However, configurations Spiral-B and Spiral-VII$^Y$ are energetically indistinguishable in 2L, because of multiple competing interactions such as the slightly reduced $A_z$, and nearly unchanged |$J_1/J_3$| ratio and $K$. The second interlayer nearest neighbor exchange parameter ($J_2^\perp$) is AFM and dominates for bulk to 2L (Tables VII and VIII in Appendix-L), resulting in interlayer AFM couplings between NiI$_2$ layers.

In the ML limit, the ratio $J_2/J_1$ changes to be negative with a larger value of −0.06 (Fig. 5c), which, together with parallel coupled biquadratic magnetic dipole interaction $B$, leads to the competed AABB-AFM and Spiral-IV$^X$ configurations. The ratios of $B$ and $J_2$ over $J_1$ determine the preferred order, that is, the larger the ratio(s) are, the more favored is the AABB-AFM configuration[24]. This is verified by the comparison between the ML-NC3 and ML-C3 structures and the enlarged ratios in the reinforced AABB-AFM state under tensile strain (Tables VII and VIII).

In summary, we exploited four multiferroic phases of NiI$_2$ from bulk to the ML limit. Their magnetic groundstates are Spiral-B, Spiral-VII$^Y$, Spiral-V$^X$, Spiral-IV$^X$, and AABB-AFM. Those magnetic spirals induce in-plane electric polarizations through the inverse D−M interaction. Thus, ML NiI$_2$ is a type-II multiferroic material with multiferroicity between AABB-AFM+AFE and Spiral-IV$^X$ + FE tunable by structural details. These fruitful variations arise from competitions among layer-dependent Heisenberg (an)isotropic exchanges, biquadratic and Kitaev interactions, and single-ion anisotropy. While anisotropic and Kitaev interactions and single-ion anisotropy play a paramount role in the bulk limit [24], the competing Heisenberg exchanges and biquadratic interaction dominate the groundstates in the ML limit. Our results highlight the importance of the layer thickness and geometry in exploring the multiferroic properties of vdW layers down to the ML limit, although the properties causing variation in magnetic interactions require further understanding.

## ACKNOWLEDGMENTS

We thank Profs. Hongxin Yang at Nanjing University, Shiwei Wu at Fudan University, and Mingxing Chen at Hunan Normal University for valuable discussions. We gratefully acknowledge financial support from the Ministry of Science and Technology (MOST) of China (Grants No. 2018YFE0202700 and 2023YFA1406500), the National Natural Science Foundation of China (Grants No. 11974422 and 12204534), the Strategic Priority Research Program of the Chinese Academy of Sciences (Grant No. XDB30000000), the Fundamental Research Funds for the Central Universities, and Research Funds of Renmin University of China (Grants No. 22XNKJ30). N.L. is grateful to the China Postdoctoral Science Foundation (2022M713447) for partial financial support. All calculations for this study were performed at the Physics Lab of High-Performance Computing (PLHPC) and the Public Computing Cloud (PCC) of Renmin University of China.

**APPENDIX A: SUPPLEMENTARY METHODS**

Density functional theory (DFT) calculations were performed using the generalized gradient approximation (GGA) for the exchange–correlation potential, the projector augmented wave (PAW) method [44, 45], and a plane-wave basis set as implemented in the Vienna *ab initio* simulation package (VASP) [46, 47]. A kinetic energy cutoff of 700 (650) eV for the plane waves was used for structural optimization (calculations on the relative energies). A vacuum layer over 20 Å in thickness in the *z* direction was adopted to eliminate interactions among image layers. The on-site Coulomb interaction was considered with a *U* value of 4.2 eV and a *J* value of 0.8 eV for Ni 3*d* orbitals, according to the literature [12, 48] and our energy test calculations of a non-collinear (NCL) and four collinear (CL) magnetic configurations (Fig. 6 in Appendix-B). Spin–orbit coupling (SOC) was considered in all total energy calculations. We used the ferromagnetic (FM) configuration as a reference for the comparison of the total energies in differently sized supercells. The total energies of the FM configuration only differ by 0.08 meV/Ni, indicating good energy convergence for our calculations. Each spiral magnetic order was modeled using a certain propagation vector *q* within the first Brillouin zone of a $1 \times \sqrt{3} \times 1$ supercell using the generalized Bloch theorem (gBT) [49]. We also constructed supercells of those spiral orders with lower energies obtained by the gBT along the Y-G-X path in the first Brillouin zone for different layers to verify their relative stabilities under consideration of SOC. The Berry phase method [50] was adopted to evaluate the spiral magnetic order-induced electric polarization. The FM state, showing no electric polarization, was used as the reference state to show the layer dependence of the electric polarization values. The transition barrier for monolayer (ML) $NiI_2$ from the in-plane antiferroelectric (AFE) phase of the AABB antiferromagnetic (AFM) order to a non-electric phase with $C_3$ symmetry was calculated using the climbing image nudged elastic band (CINEB) method [51]. All atoms, lattice volumes, and shapes in each supercell were allowed to relax until the residual force on each atom was less than 0.01 eV/Å. Grimme's semiempirical D3 scheme [52] for dispersion correction was employed to describe the vdW interactions in combination with the Perdew–Burke–Ernzerhof functional (PBE-D3) [53]. This combination achieved accuracy comparable to that of the optB86b-vdW functional for describing geometric properties of layered materials [54] at a lower computational cost.

The nearest-neighbor exchange-coupling tensor for Ni–Ni pairs in the *xyz* basis is:

$$J_1^{xyz} = \begin{pmatrix} J_{xx} & 0 & 0 \\ 0 & J_{yy} & J_{yz} \\ 0 & J_{yz} & J_{zz} \end{pmatrix}. \tag{A1}$$

If $J_{yz} \neq 0$, adjacent moments will be non-collinearly coupled and lie in a plane off the *xy*-plane [29].

The tensor $J_1^{xyz}$ can be diagonalized in the *αβγ* basis shown in Fig. 5a in the main text as

$$J_1^{\alpha\beta\gamma} = \begin{pmatrix} \lambda_\alpha & 0 & 0 \\ 0 & \lambda_\beta & 0 \\ 0 & 0 & \lambda_\gamma \end{pmatrix}. \tag{A2}$$

Therefore, the nearest-neighbor exchange-coupling Hamiltonian in such a basis can be written as

$$H_1 = -\frac{1}{2}\sum_{i \neq j}(\lambda_\alpha S_i^\alpha S_j^\alpha + \lambda_\beta S_i^\beta S_j^\beta + \lambda_\gamma S_i^\gamma S_j^\gamma). \tag{A3}$$

For NiI$_2$, we assume $\lambda_\alpha = \lambda_\beta$. $H_1$ then can be expressed as

$$H_1 = -\frac{1}{2}\sum_{i \neq j}(J\mathbf{S}_i \cdot \mathbf{S}_j + KS_i^\gamma S_j^\gamma), \tag{A4}$$

where $J = (\lambda_\alpha + \lambda_\beta)/2$ is the isotropic nearest-exchange coupling in the *αβ*-plane and $K = \lambda_\alpha - J$ is the Kitaev anisotropic nearest-exchange coupling parameter.

The calculated $J_1^{xyz}$ of ML NiI$_2$ with the C$_3$ symmetry is

$$J_1^{xyz} = \begin{pmatrix} 3.67 & 0 & 0 \\ 0 & 4.64 & 0.20 \\ 0 & 0.20 & 3.68 \end{pmatrix} \tag{A5}$$

and thus

$$J_1^{\alpha\beta\gamma} = \begin{pmatrix} 3.74 & 0 & 0 \\ 0 & 3.74 & 0 \\ 0 & 0 & 4.51 \end{pmatrix}. \tag{A6}$$

Therefore, $J_{yz}$ = 0.20 meV, $J$ = 3.74 meV, and $K$ = 0.77 meV. The $J_{yz}$, $J$, and $K$

parameters for the other NiI$_2$ layers can be obtained in a similar way.

# APPENDIX B: ENERGY COMPARISONS FOR BULK NiI$_2$

We considered 5 collinear (Fig. 6) and 55 NCL (Figs. 1e and 1f) magnetic configurations. The experimentally observed spiral order (Spiral-B) shows the lowest energy among these magnetic orders (Fig. 1f and Table I), which is robust regardless of the preservation of the C$_3$ symmetry (Fig. 1e), consideration of SOC (Fig. 7), and choice of on-site Coulomb interaction (Fig. 8) values.

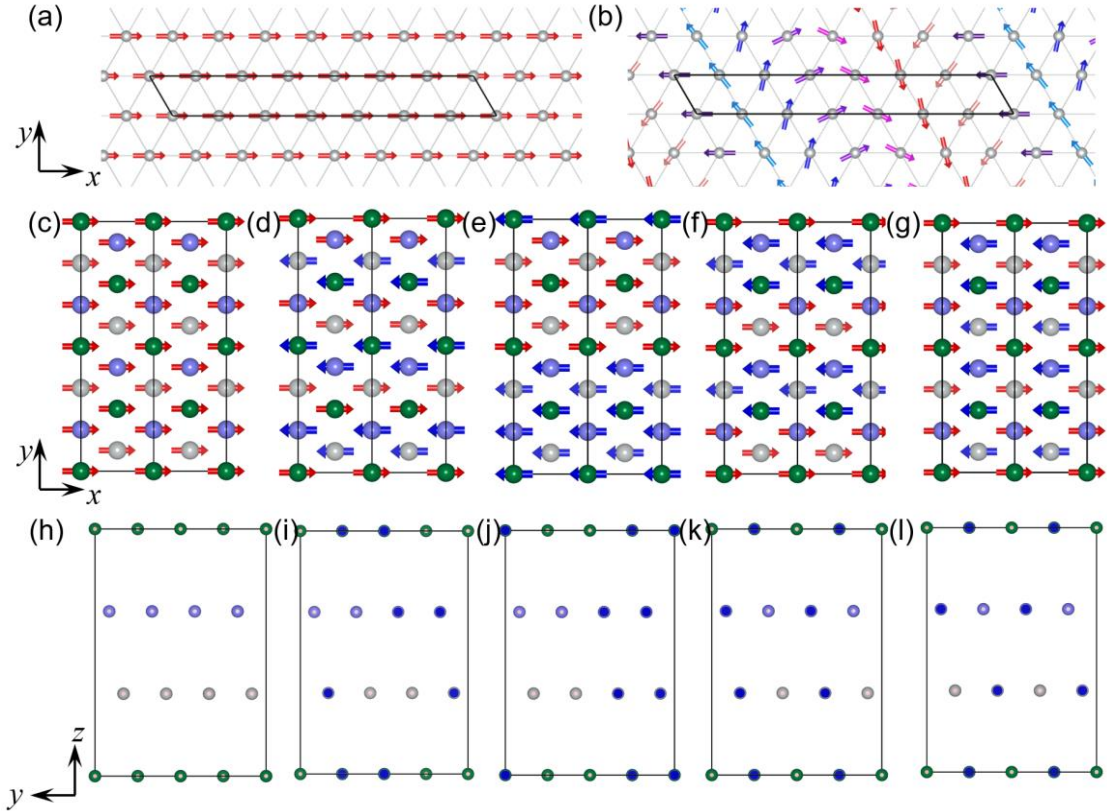

FIG. 6. Schematic models of various collinear and non-collinear (NCL) magnetic configurations for bulk NiI$_2$ considered in this work, including top views of (a) FM and (b) spiral orders in a $7 \times 1 \times 1$ supercell, and (c) FM, (d) AABB-ABBA-AABB-AFM, (e) AABB-AABB-AABB-AFM, (f) ABAB-ABAB-ABAB-AFM, and (g) AABB-BABA-ABAB-AFM orders in a $1 \times 2\sqrt{3} \times 1$ supercell, with magnetic moments along the $x$ direction. The configuration "AABB-ABBA-AABB-AFM" means AABB-AFM, ABBA-AFM and AABB-AFM magnetic orders in the 1$^{st}$, 2$^{nd}$, and 3$^{rd}$ NiI$_2$ layers, which are labeled by green, grey, and blue balls, respectively. (h-l)

Side views corresponding to (c-g). The black boxes represent different supercells. Directions of magnetic moments were labeled by colored arrows.

TABLE I. Relative energies ($\Delta E$) of bulk NiI$_2$ in various collinear and NCL magnetic configurations (shown in Fig. 6). The energy of the Spiral-B [$q_B$ = (0, 0.138, 1.457)] order was set to the reference zero. The values in parentheses represent the cases without the C$_3$ symmetry. Lattice m × n indicates the number of supercells in the $a$ and $b$ directions. Spiral-B is the magnetic groundstate of bulk NiI$_2$.

| Lattice | Mag. Config. | $\Delta E$ (meV/Ni) |
|---|---|---|
| 4 × √3 | $q_{IV}^X$ = (0.250, 0, 0) | 0.64 (0.64) |
| 5 × √3 | $q_V^X$ = (0.200, 0, 0) | 0.19 |
| 1 × 4 | $q_{IV}^Y$ = (0, 0.250, 0) | 1.91 |
| 1 × 5 | $q_V^Y$ = (0, 0.200, 0) | 0.31 |
| 1 × 6 | $q_{VI}^Y$ = (0, 0.167, 0) | 0.21 |
| 1 × 7 | FM | 10.82 |
| 1 × 7 | $q_{VII}^Y$ = (0, 0.143, 0) | 0.05 |
| 1 × 7 | $q_B$ = (0, 0.138, 1.457) | 0 |
| 1 × 8 | $q_{VIII}^Y$ = (0, 0.125, 0) | 0.48 |
| 1 × 2√3 | AABB-AABB-AABB-AFM | 1.43 (0.73) |
| 1 × 2√3 | AABB-ABBA-BBAA-AFM | 1.57 |
| 1 × 2√3 | ABAB-ABAB-ABAB-AFM | 18.33 |
| 1 × 2√3 | ABAB-BABA-ABAB-AFM | 23.17 |

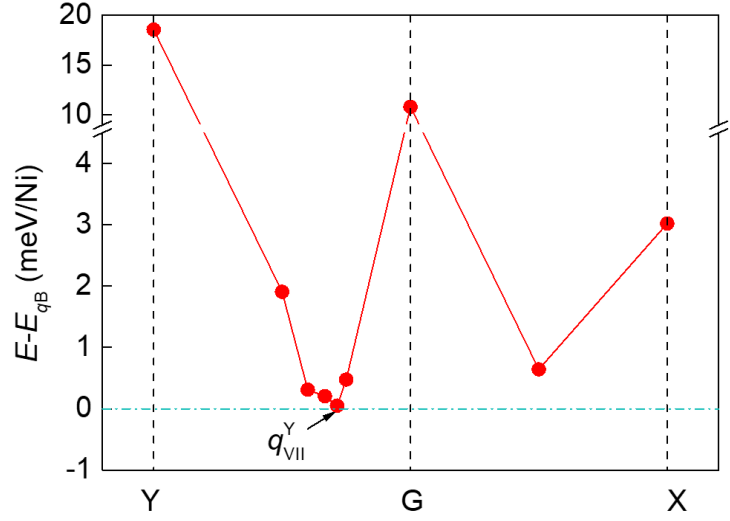

FIG. 7. Energies of various spiral orders along the Y-G-X path in different supercells relative to the total energy of the $q_B$ order (represented by the horizonal dashed cyan line) in bulk NiI$_2$. Spin-orbit coupling was considered in all supercells. The comparable order $q_{VII}^Y$ is labeled, which is the in-plane projection of Spiral-B. Spiral-B is the groundstate of bulk NiI$_2$.

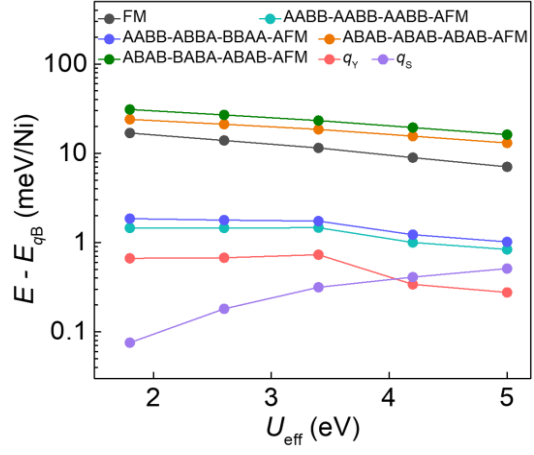

FIG. 8. Effect of different on-site effective Coulomb $U$ values ($U_{\text{eff}}$) on bulk $NiI_2$. Comparison of total energies of different magnetic orders relative to that of $q_B$ as a function of $U_{\text{eff}}$ values using PBE-D3 functional. The two additional spiral orders are $q_Y = (0, 0.5, 0)$ and $q_S = (0, 0.5, 0.5)$. The Spiral-B ground state of bulk $NiI_2$ is robust regardless of the choice of $U_{\text{eff}}$ values.

# APPENDIX C: TOTAL ENERGY COMPARISONS OF VARIOUS MAGNETIC CONFIGURATIONS IN MONOLAYER NiI$_2$

For monolayer NiI$_2$, we considered 3 collinear and 19 NCL magnetic orders (Figs. 9 to 13 and Table II) in different supercell sizes (Fig. 9) and a $10 \times 10$ $q$-mesh (Fig. 2a) for the gBT model to explore the magnetic groundstate. Different from the Spiral-B in bulk NiI$_2$, a new spiral order (Spiral-IV$^X$) emerges in ML NiI$_2$ with the C$_3$ symmetry (ML-C3). The magnetic moments of Spiral-IV$^X$ align in the Ni-I plane (Fig. 10b) and follow a propagating vector $q_{IV}^X = (0.250, 0, 0)$ (across each $4 \times \sqrt{3}$ supercell, Figs. 10a). For the ML NiI$_2$ without the C$_3$ symmetry (ML-NC3), the AABB-AFM state is more stable than Spiral-IV$^X$ by 0.35 meV/Ni (Table II). The magnetic groundstate of ML NiI$_2$ is thus dependent on the structural symmetry, indicating that external strains play a role in modulating their relative stability.

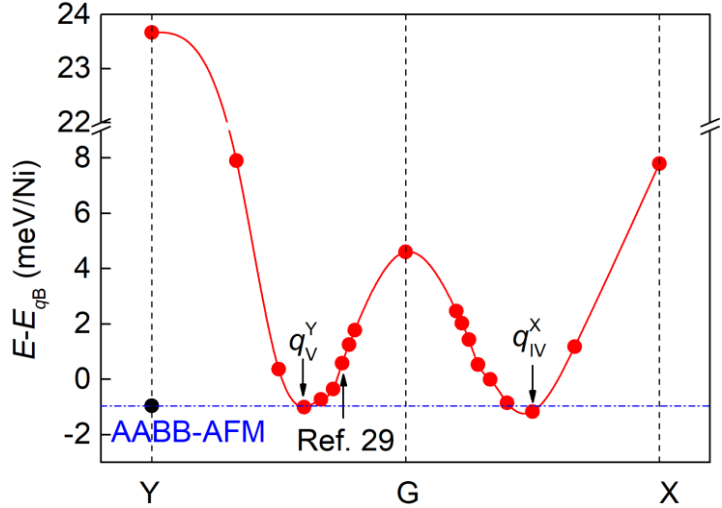

FIG. 9. Comparison of total energies among various spiral orders in different supercells for ML NiI$_2$ with the C$_3$ symmetry (ML-C3). The energy of the $q_B$ order was set to the reference zero. Spin-orbit coupling was considered in all supercells. The horizonal dashed blue line represents the energy of AABB-AFM order. The energy of ML NiI$_2$ in Ref. [29] is also indicated.

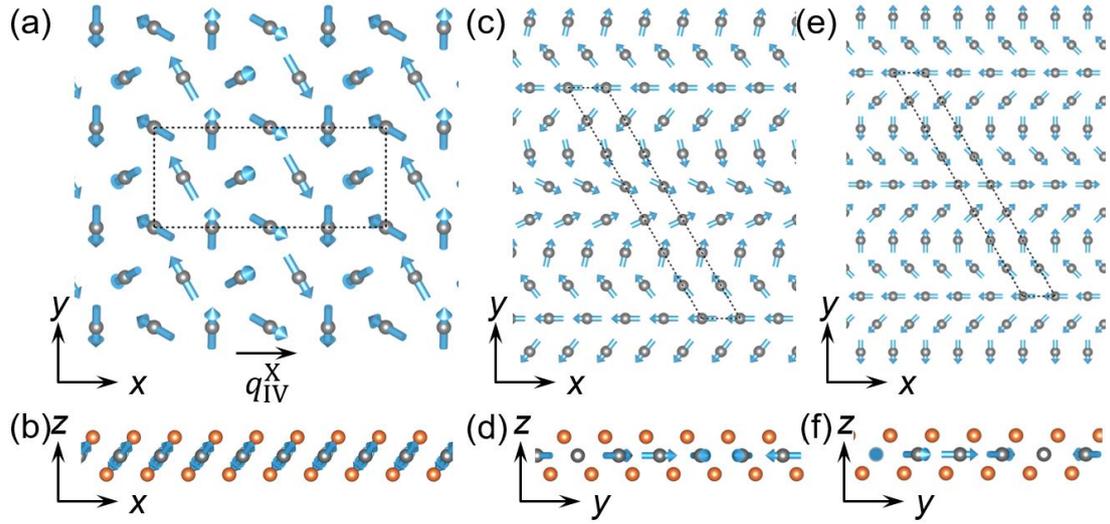

FIG. 10 Schematic plots of spiral orders of ML NiI$_2$. (a, b) Top and side views of $q_{IV}^{X}$ with magnetic moments lying in the $\alpha\beta$-plane (illustrated in Fig. 5a). Spiral orders with propagation vectors along the *y* direction across (c, d) a $1 \times 7$ and (e, f) a $1 \times 8$ supercell, the magnetic moments of which lie in the *xy*-plane.

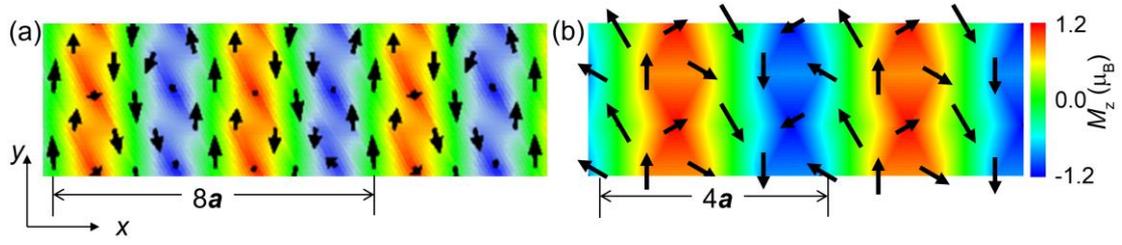

FIG. 11. Magnetic configurations of (a) the spiral order in Fig. 3c of Ref. [11] and (b) Spiral-IV$^X$. In both figures, black arrows indicate the in-plane components of magnetic moments, while the color maps represent the out-of-plane components. Their propagating vectors of 8$a$ and 4$a$ were labeled in panels (a) and (b), respectively.

TABLE II. Relative energies $\Delta E$ of ML NiI$_2$ with the C$_3$ symmetry limitation (ML-C3) in different collinear and NCL magnetic configurations. The energy of $q_B$ order was set to the reference zero. The values in parentheses represent the cases without the C$_3$ symmetry (ML-NC3). The magnetic ground state of ML-C3 is Spiral-IV$^X$ ($q_{IV}^X$), while that for ML-NC3 is AABB-AFM order.

| Lattice | Mag. Config. | $\Delta E$ (meV/Ni) |
|---|---|---|
| 2×√3 | $q_{II}^X$ = (0.500, 0, 0) | 7.80 (7.34) |
| 3×√3 | $q_{IV}^X$ = (0.333, 0, 0) | 1.19 (0.90) |
| 4×√3 | $q_{IV}^X$ = (0.250, 0, 0) | −1.16 (−1.33) |
| 5×√3 | $q_V^X$ = (0.200, 0, 0) | −0.84 (−0.93) |
| 6×√3 | $q_{VI}^X$ = (0.167, 0, 0) | 0.01 (−0.05) |
| 7×√3 | $q_{VII}^X$ = (0.143, 0, 0) | 0.54 (0.51) |
| 8×√3 | $q_{VIII}^X$ = (0.125, 0, 0) | 1.44 (1.42) |
| 1×2 | $q_{II}^Y$ = (0, 0.500, 0) | 23.67 (19.50) |
| 1×3 | $q_{III}^Y$ = (0, 0.333, 0) | 7.91 (8.04) |
| 1×4 | $q_{IV}^Y$ = (0, 0.250, 0) | −0.37 (0.38) |
| 1×5 | $q_V^Y$ = (0, 0.200, 0) | −1.00 (−1.04) |
| 1×6 | $q_{VI}^Y$ = (0, 0.167, 0) | −0.72 (−0.77) |
| | FM | 4.61 (5.37) |
| 1×7 | $q_{VII}^Y$ = (0, 0.143, 0) | −0.35 (−0.38) |
| | $q_B$ = (0, 0.138, 1.457) | 0 (−0.15) |
| 1×8 | $q_{VIII}^Y$ = (0, 0.125, 0) | 0.59 (0.55) |
| 3×2 | NCL | 15.23 |
| 3×3 | NCL | 8.35 |
| √3×√3 | NCL | 8.82 |
| 2√3×√3 | NCL | 4.10 |
| 1×2√3 | AABB-AFM | −0.95 (−1.68) |
| | ABAB-AFM | 19.46 (24.17) |

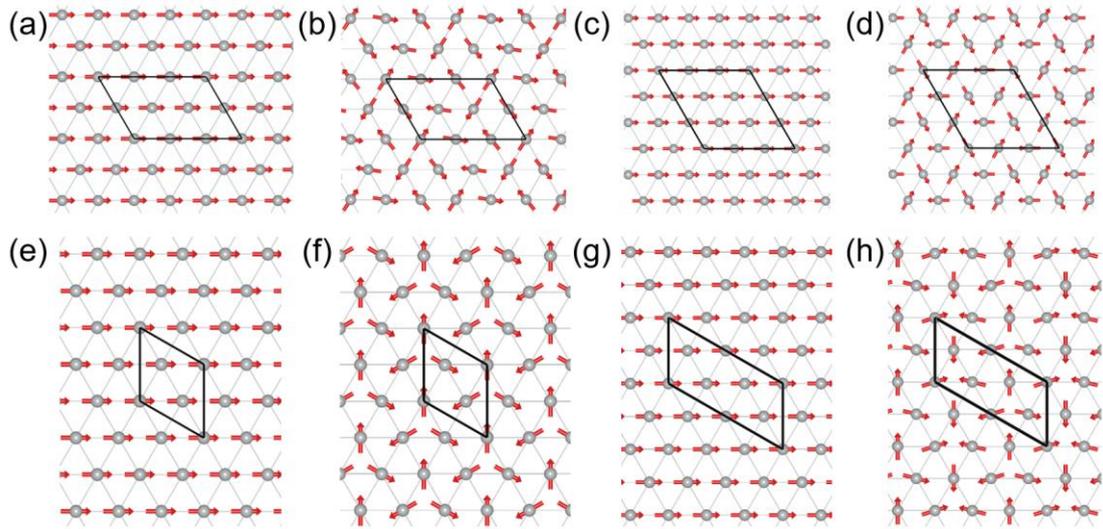

FIG. 12. Schematic plots of collinear and NCL magnetic configurations for ML NiI$_2$ considered in this work, including (a) FM and (b) NCL in a 3 × 2 supercell, (c) FM and (d) NCL in a 3 × 3 supercell, (e) FM and (f) NCL in a √3 × √3 supercell, and (g) FM and (h) NCL in a 2√3 × √3 supercell. The black boxes represent different supercells. Directions of magnetic moments were indicated by red arrows.

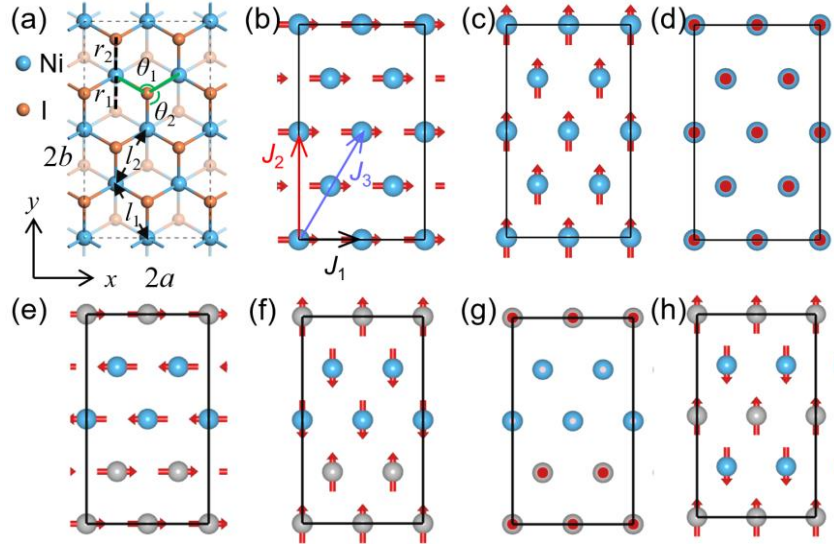

FIG. 13. (a) Structural details of ML NiI$_2$. The black rectangle illustrates the $2 \times 2\sqrt{3}$ supercell. The blue and orange spheres indicate Ni and I atoms, respectively. Several key structural parameters, including $r_1$, $r_2$, $\theta_1$, $\theta_2$, $l_1$, and $l_2$, were marked. Different collinear magnetic configurations with magnetic moments along the *x*, *y*, and *z* directions were shown, including (b) FM order along the *x* direction, (c) along the *y* direction, (d) along the *z* direction, (e) AABB-AFM order along the *x* direction, (f) along the *y* direction, (g) along the *z* direction, and (h) ABAB-AFM order along the *y* direction.

# APPENDIX D: STRAIN EFFECTS ON THE MAGNETIC GROUNDSTATE OF ML NiI$_2$

We also carried out calculations to compare the stability of some competing magnetic configurations under strains. Configurations $q_V^Y$ and $q_{VII}^Y$ were compared with $q_{IV}^X$ and AABB-AFM under *x-y* epitaxial strains for NiI$_2$ MLs without (ML-NC3) and the AABB-AFM configuration was also examined under biaxial strains for ML NiI$_2$ with the C$_3$ symmetry (ML-C3). Figure 14a clearly illustrates that either the AABB-AFM or the Spiral-IV$^X$ order exhibits the best stability among all four considered magnetic orders. In particular, Spiral-IV$^X$ (red plane) is always more stable than $q_V^Y$ (by 0.12 to 0.41 meV/Ni, yellow plane in Fig. 14a), $q_{VII}^Y$ (by 0.42 to 1.67 meV/Ni, green plane in Fig. 14a) in ML-NC3, and AABB-AFM in ML-C3 (AABB-C3, by 0.14 to 0.31 meV/Ni, Fig. 14b) in the whole considered strain range. In certain strain regions, the AABB-AFM order (blue plane in Fig. 14a) becomes more stable than Spiral-IV$^X$. Based on these results, we verified the reliability of only comparing the energies of $q_{IV}^X$ and AABB-AFM for competing the magnetic ground state of ML NiI$_2$.

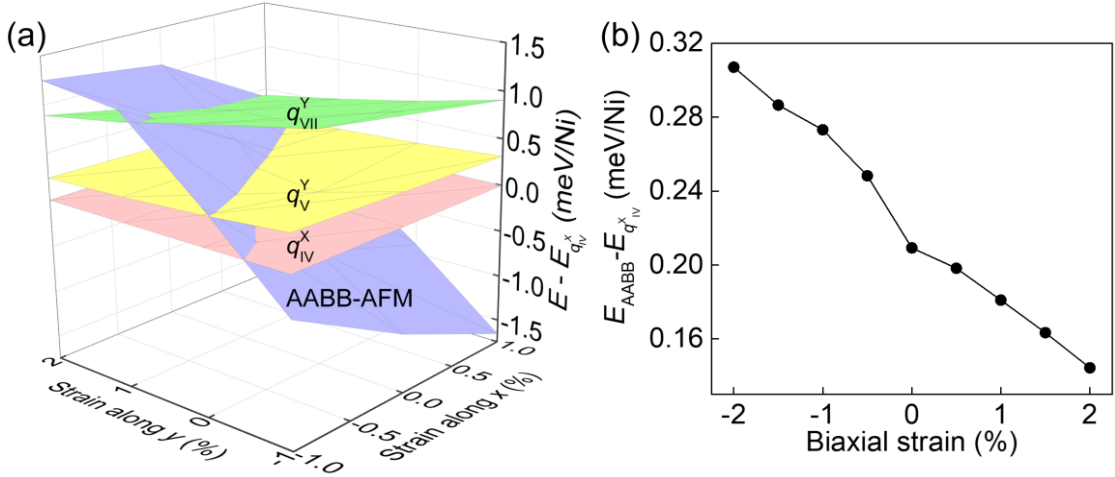

FIG. 14. (a) Relative energies of Spiral-V$^Y$ ($q_V^Y$), Spiral-VII$^Y$ ($q_{VII}^Y$), and AABB-AFM to that of Spiral-IV$^X$ ($q_{IV}^X$) for ML-NC3 under epitaxial strains along the *x* and *y* directions. (b) Relative energy of AABB-AFM to $q_{IV}^X$ for ML-C3 as a function of biaxial strains.

# APPENDIX E: LAYER-DEPENDENT MAGNETIC GROUNDSTATES FOR FEWLAYERS NiI$_2$

For constrain-free 2L NiI$_2$, the AABB-AFM order is less stable than other NCL orders (Table III). A new spiral order, named Spiral-V$^X$ emerges and is 0.26 meV/Ni more stable than Spiral-IV$^X$. The magnetic moments of Spiral-V$^X$ propagate along the $x$ direction and across each $5 \times \sqrt{3}$ supercell. Spiral-V$^X$ is also more stable than Spiral-VII$^Y$ and Spiral-B by 0.24 and 0.18 meV/Ni, nearly energetically undistinguishable. The Spiral-VII$^Y$ is the in-plane projection of Spiral-B and propagates along the $y$ direction across a $1 \times 7$ supercell. Breakdown or preservation of the C$_3$ symmetry doesn't change the relative stability of these spiral orders (Table III). For 3L NiI$_2$, Spiral-VII$^Y$ and Spiral-B are still energetically degenerated, but over 0.35 meV/Ni more stable than Spiral-IV$^X$ and Spiral-V$^X$ (TABLE IV). For 4L and thicker NiI$_2$, Spiral-B becomes the most stable among various collinear and NCL magnetic configurations (Table V).

TABLE III. Relative energies $\Delta E$ of bilayer (2L) NiI$_2$ with the C$_3$ symmetry limitation (2L-C3) in different collinear and NCL magnetic configurations. The energy of $q_B$ order was set to reference zero. The values in parentheses represent the cases without the C$_3$ symmetry (2L-NC3).

| Lattice | Mag. Config. | $\Delta E$ (meV/Ni) |
|---|---|---|
| $4 \times \sqrt{3}$ | $q_{IV}^X = (0.250, 0, 0)$ | 0.08 (0.02) |
| $5 \times \sqrt{3}$ | $q_V^X = (0.200, 0, 0)$ | −0.18 (−0.08) |
| $6 \times \sqrt{3}$ | $q_{VI}^X = (0.167, 0, 0)$ | 0.27 (0.28) |
| $7 \times \sqrt{3}$ | $q_{VII}^X = (0.143, 0, 0)$ | 0.42 (0.39) |
| $1 \times 4$ | $q_{IV}^Y = (0, 0.250, 0)$ | 1.63 |
| $1 \times 5$ | $q_V^Y = (0, 0.200, 0)$ | −0.03 (0.14) |
| $1 \times 6$ | $q_{VI}^Y = (0, 0.167, 0)$ | 0.11 |
| $1 \times 7$ | FM | 8.11 (8.48) |
| | $q_{VII}^{Y} = (0, 0.143, 0)$ | 0.06 (−0.01) |
| | $q_{B} = (0, 0.138, 1.457)$ | 0 (−0.02) |
| 1×8 | $q_{VIII}^{Y} = (0, 0.125, 0)$ | 0.64 |
| 1×2√3 | AABB-AABB-AFM | 0.56 (0.27) |
| | AABB-ABBA-AFM | 1.03 (0.62) |

TABLE IV. Relative energies $\Delta E$ of trilayer (3L) NiI$_2$ with the C$_3$ symmetry limitation (3L-C3) in different collinear and NCL magnetic configurations. The energy of $q_B$ order was set to reference zero. The values in parentheses represent the cases without the C$_3$ symmetry (3L-NC3).

| Lattice | Mag. Config. | $\Delta E$ (meV/Ni) |
|---|---|---|
| 4×√3 | $q_{IV}^X$ = (0.250, 0, 0) | 0.65 (0.90) |
| 5×√3 | $q_V^X$ = (0.200, 0, 0) | 0.35 |
| 6×√3 | $q_{VI}^X$ = (0.167, 0, 0) | 0.50 |
| 1×4 | $q_{IV}^Y$ = (0, 0.250, 0) | 2.19 |
| 1×5 | $q_V^Y$ = (0, 0.200, 0) | 0.44 |
| 1×6 | $q_{VI}^Y$ = (0, 0.167, 0) | 0.20 |
| 1×7 | FM | 9.44 |
| 1×7 | $q_{VII}^Y$ = (0, 0.143, 0) | 0.06 |
| 1×7 | $q_B$ = (0, 0.138, 1.457) | 0 |
| 1×8 | $q_{VIII}^Y$ = (0, 0.125, 0) | 0.61 |
| 1×2√3 | AABB-AABB-AABB-AFM | 1.17 (0.45) |
| 1×2√3 | AABB-ABBA-BBAA-AFM | 1.84 |
| 1×2√3 | ABAB-ABAB-ABAB-AFM | 18.01 |
| 1×2√3 | ABAB-BABA-ABAB-AFM | 22.76 |

TABLE V. Relative energies $\Delta E$ of four-layer (4L) NiI$_2$ with the C$_3$ symmetry limitation (4L-C3) in different collinear and NCL magnetic configurations. The energy of $q_B$ order was set to reference zero. The values in parentheses represent the cases without the C$_3$ symmetry (4L-NC3).

| Lattice | Mag. Config. | $\Delta E$ (meV/Ni) |
|---|---|---|
| 4×√3 | $q_{IV}^X = (0.250, 0, 0)$ | 0.86 (0.70) |
| 5×√3 | $q_V^X = (0.200, 0, 0)$ | 0.52 |
| 6×√3 | $q_{VI}^X = (0.167, 0, 0)$ | 0.57 |
| 1×4 | $q_{IV}^Y = (0, 0.250, 0)$ | 2.40 |
| 1×5 | $q_V^Y = (0, 0.200, 0)$ | 1.28 |
| 1×6 | $q_{VI}^Y = (0, 0.167, 0)$ | 0.30 |
| 1×7 | FM | 10.15 |
| | $q_{VII}^Y = (0, 0.143, 0)$ | 0.14 |
| | $q_B = (0, 0.138, 1.457)$ | 0 |
| 1×8 | $q_{VIII} = (0, 0.125, 0)$ | 0.64 |
| 1×2√3 | AABB-AABB-AABB-AABB-AFM | 1.44 (0.70) |
| | AABB-AABB-ABBA-ABBA-AFM | 1.69 |
| | AABB-ABBA-ABBA-AABB-AFM | 1.55 |
| | AABB-AABB-AABB-ABBA-AFM | 1.69 |
| | ABBA-ABBA-ABBA-AABB-AFM | 1.31 |

# APPENDIX F: EFFECTS OF DIFFERENT $U_{eff}$ VALUES AND FUNCTIONALS ON THE MAGNETIC STABILITY OF ML NiI$_2$

We examined the effect of on-site effect $U_{eff}$ values and functionals on the magnetic groundstate of ML NiI$_2$. As shown in Fig. 15a, for ML-NC3 (ML-C3), the AABB-AFM ($q_{IV}^X$) state always shows the lowest energy under different $U_{eff}$ values and functionals (orange squares). The relative stability for these magnetic configurations was also checked using the HSE06 functional (Fig. 15b).

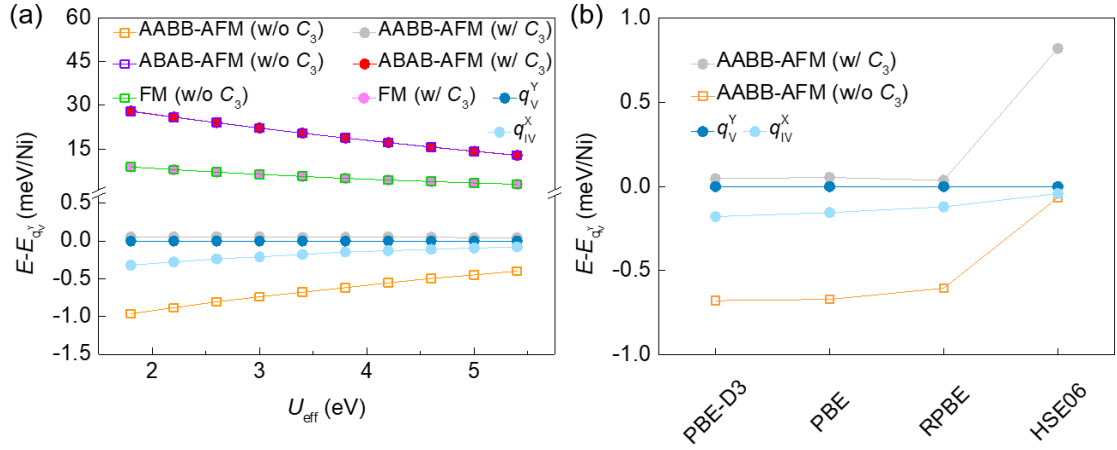

FIG. 15. Total energies of different magnetic orders to that of the Spiral-V$^Y$ state were shown as functions of (a) effective $U_{eff}$ values using PBE-D3 functional and (b) functionals with $U_{eff}$ = 3.8 eV. Spin-orbit coupling was considered in the PBE-D3, PBE, and RPBE functionals. $U_{eff}$ was not used for the HSE06 functional calculation.

# APPENDIX G: ELECTRIC PROPERTIES FOR ML NiI$_2$ WITH AND WITHOUT THE C$_3$ SYMMETRY

We used the NEB method to calculate the energy barriers for the spin canting process of ML-C3 from $q_{IV}^X$ to $-q_{IV}^X$ (Fig. 16a) and AFE transition of ML-NC3 through a non-electric phase (Fig. 16b). As listed in Table VI, the lattice *b* shrinks for the AABB-AFM state relative to that of FM order. In the AABB-AFM order, the Ni atoms in the same AFM stripe move oppositely by 0.004 Å in the ±*y* directions, yielding a dipole moment of ±1.5 pC/m (Fig. 3g).

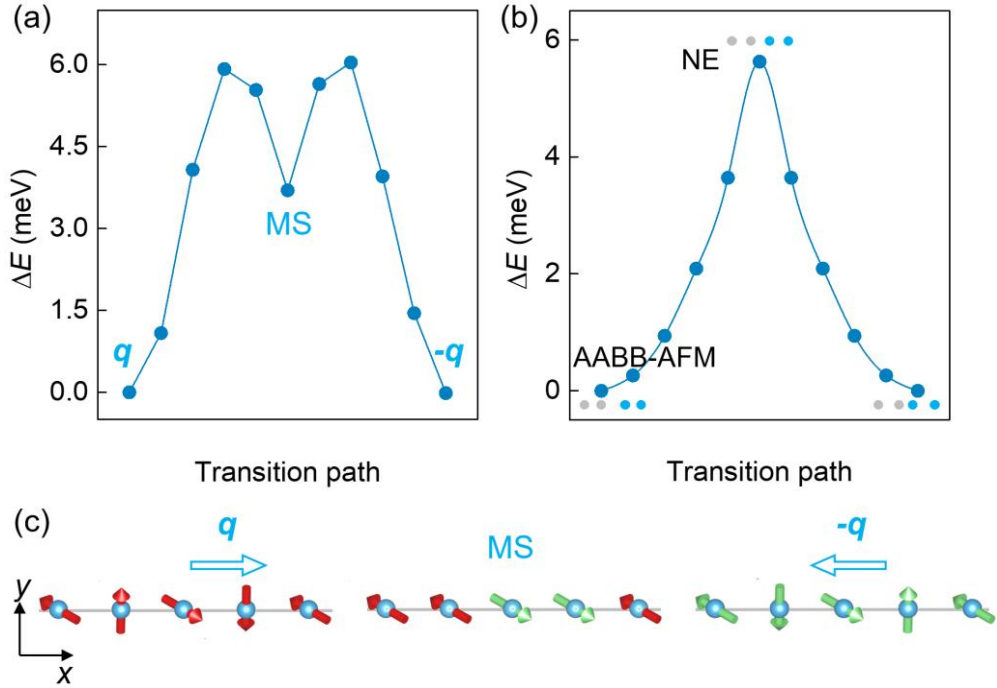

FIG. 16. (a) Calculated energy barrier for the spin canting process of ML NiI$_2$, taking Spiral-IV$^X$ order (illustrated in Figs. 3a-3c). (b) Transition pathway for ML NiI$_2$ between the two AFE phases through a non-electric (NE) phase. The insets blue and gray circles represent the Ni atoms in different magnetic stripes. (c) Initial, metastable (MS) and final magnetic configurations for the spin canting process in (a).

TABLE VI. Structural details of ML NiI$_2$ in a $2 \times 2\sqrt{3}$ supercell with different magnetic configurations (marked in Fig. 13a). Sym. represents the space point group symmetry.

| Spin config. | $2a$ (Å) | $2\sqrt{3}b$ (Å) | $l_1$ (Å) | $l_2$ (Å) | $\theta_1$ (°) | $\theta_2$ (°) | Sym. |
|---|---|---|---|---|---|---|---|
| FM | 7.86 | 13.61 | 3.93 | 3.93 | 91.21 | 91.21 | P$\bar{3}$m1 |
| AABB-AFM | 7.85 | 13.56 | 3.91 | 3.93 | 90.98 | 91.15 | P2$_1$/$m$ |
| ABAB-AFM | 7.86 | 13.61 | 3.93 | 3.93 | 91.18 | 91.18 | P2$_1$/$m$ |
| ZZ-AFM | 7.87 | 13.63 | 3.91 | 3.93 | 90.95 | 91.13 | P2/C |

# APPENDIX H: SUBSTRATE EFFECT ON MAGNETIC GROUNDSTATE OF ML NiI$_2$

The relation between lattices of the ML NiI$_2$ and hBN substrate is yet to be uncovered in experiments. Without such experimental input, we theoretically constructed the superlattices of their heterostructures using a criterion of optimal lattice mismatching. We adopted the supercell with lattice of –0.6% along the $x$ direction and 1.3% along the $y$ direction (referred to as Str.1), and an alternative hexangular supercell (referred to as Str.2, Fig. 17a) and a rectangular one (referred to as Str.3, Fig. 17b), exhibiting the second and the third smallest lattices mismatches, namely 2.1% along both the $x$ and $y$ directions in Str.2, and 3.8% along the $x$ directions and –4.0% along the $y$ direction in Str.3. Please be aware that the strains in Str.2 are in the tensile region, while the $x(y)$-strain in Str.3 is in the opposite direction of that for Str.1, namely tensile (compressive) $x(y)$-strain. Our calculations indicate that Str.1 is 1.43 and 2.48 meV/Ni more stable than Str.2 and Str.3, respectively, given a defined interfacial binding energy $E_b = E_{tot} - E_{hBN} - E_{NiI2}$, where $E_{tot}$, $E_{hBN}$ and $E_{NiI2}$ denote the energies of the heterostructure, $h$BN and NiI$_2$, respectively.

In Str.1, the Ni layer (Fig. 18a) and the two I sublayers (Fig. 19) exhibit out-of-plane corrugations varying up to 0.06 Å. In-plane strains break the inversion symmetry in the $x$ (Fig. 18c) and $y$ (Fig. 18d) directions. Moreover, explicit interfacial charge transfer from the hBN substrate to the interfacial I layer leads to an out-of-plane electric polarization (Figs. 18e and 18f).

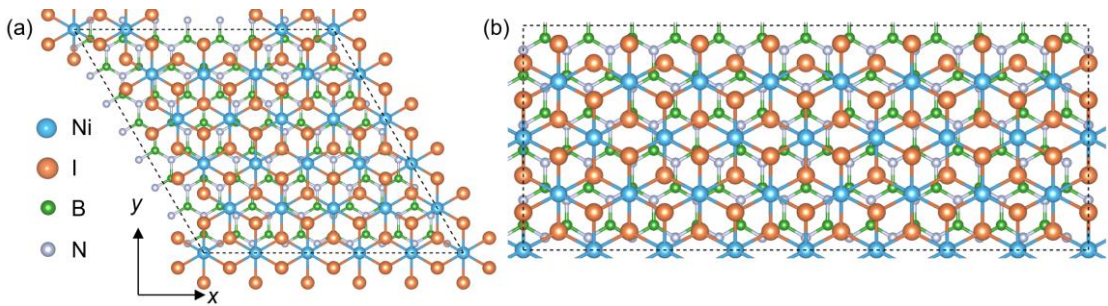

FIG. 17. Top view of two alternative NiI$_2$/hBN heterostructures in the tensile strain region, exhibiting lattice mismatches of 2.1% for the $x$ and $y$ directions in (a) and 3.8%

along the $x$ direction and –4.0% along the $y$ direction in (b). The hexangular and rectangular supercells are depicted, highlighting the relative arrangement of the layers.

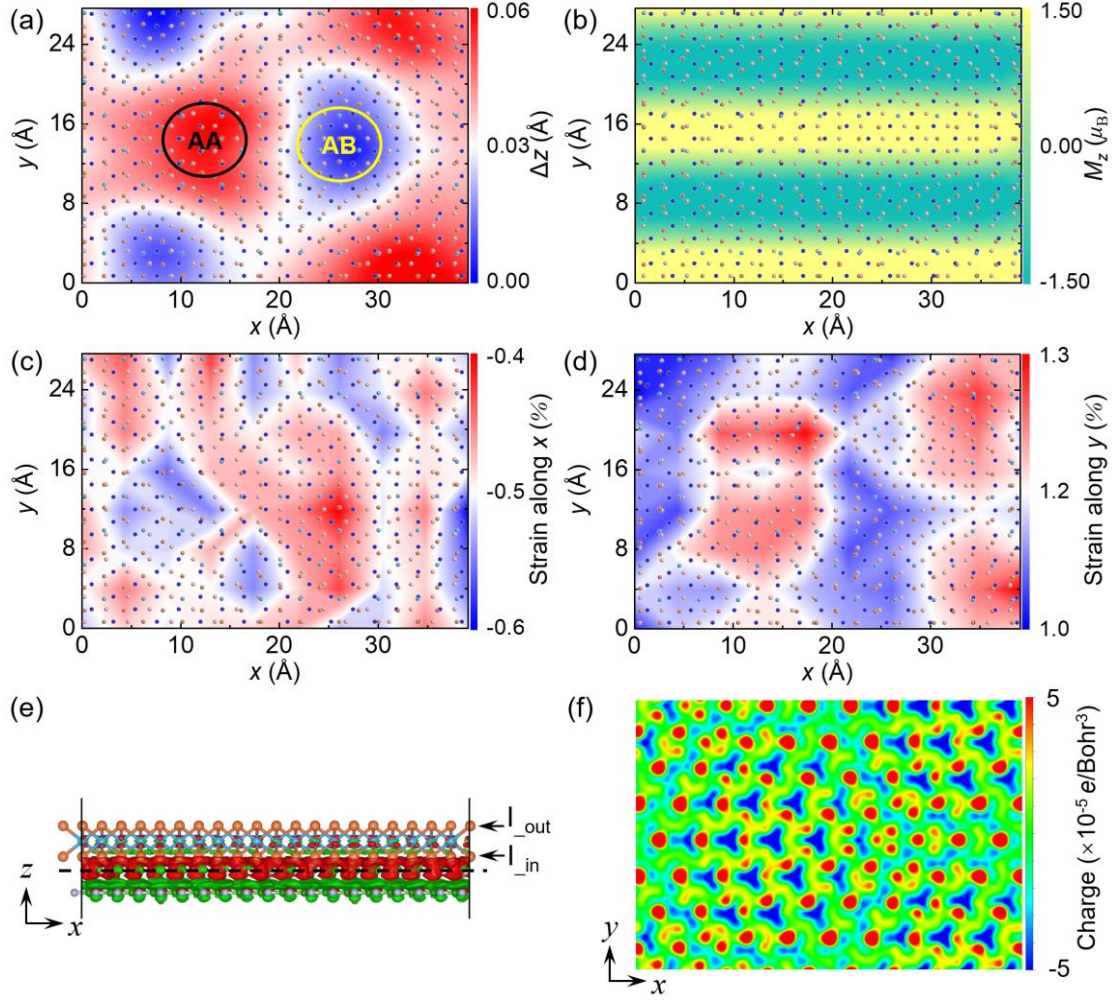

FIG. 18. Structure and magnetism of the epitaxial ML $NiI_2$ on a $h$BN substrate, including 2D mappings of the variation in (a) the $z$ coordinates and (b) the magnetic moments of Ni atoms in the heterostructure taking the AABB-AFM state, and in-plane strains along the (c) $x$ and (d) $y$ directions. (e) Side view of differential charge density and (f) 2D profile maps along the dashed lines in (e) with an isosurface value of $5 \times 10^{-5}$ $e$/Bohr$^3$. The red and green contours in (e) and (f) are charge accumulation and depletion, respectively. The lower and upper I atoms of ML $NiI_2$ on $h$BN are labeled by $I_{in}$ and $I_{out}$ in (e).

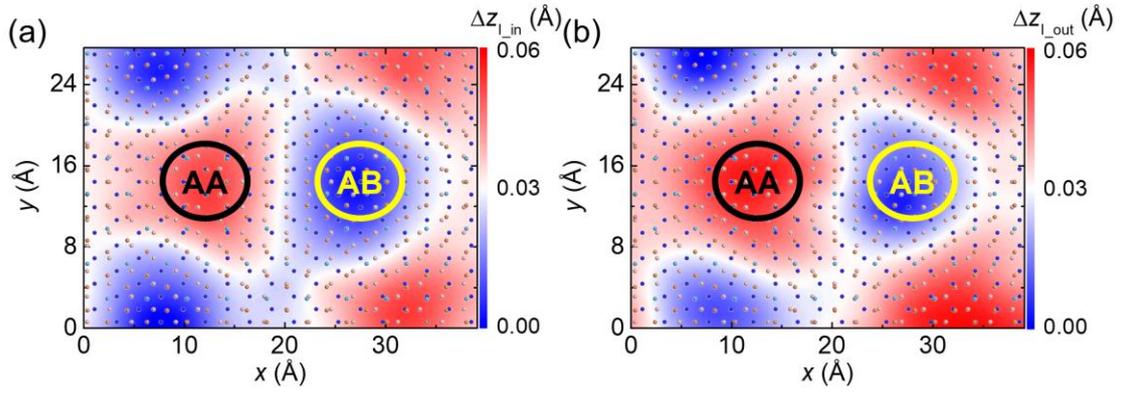

FIG. 19. Two-dimensional mappings of the $z$-coordinates of the (a) $I_{\_in}$ and (b) $I_{\_out}$ atoms (labeled in Fig. 18e) of ML $NiI_2$ adsorbed on the hBN substrate. The spatial variations for $z$-coordinates of the $I_{\_in}$ and $I_{\_out}$ indicate out-of-plane inversion symmetry breaking for $NiI_2$ on hBN substrate.

# APPENDIX I: EFFECT OF ELECTRON DOPING ON THE RELATIVE STABILITY BETWEEN SPIRAL-IV$^X$ AND AABB-AFM ORDERS

We considered the effect of electron doping on the relative stability between AABB-AFM and Spiral-IV$^X$ for the both structures. As shown in Fig. 20, for ML-C3, the Spiral-IV$^X$ is always more stable than the AABB-AFM (red dots), while it is preferred for doping concentration larger than 0.02 $e$/I for ML-NC3 (black dots). The hBN substrate interacts strongly with ML NiI$_2$, the C$_3$ symmetry of ML NiI$_2$ thus is, most likely, maintained on hBN. Moreover, electrons are transferred from hBN to ML NiI$_2$ (Fig. 18e). The magnetic groundstate of ML NiI$_2$ on hBN is thus the Spiral-IV$^X$.

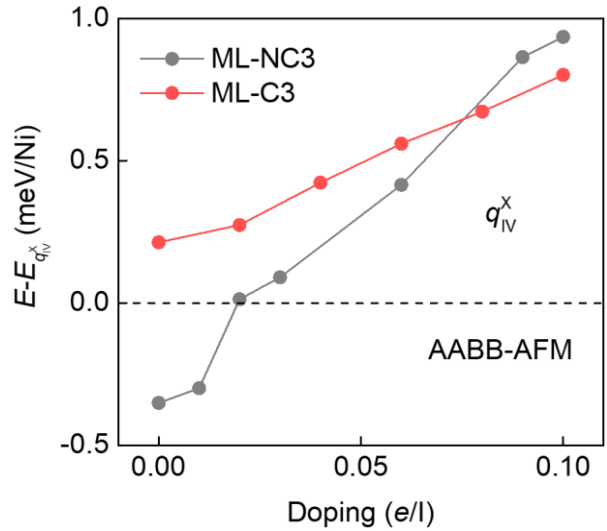

FIG. 20. Energy difference between AABB-AFM and Spiral-IV$^X$ orders for ML-NC3 (black) and ML-C3 (red) under different doping concentrations. For ML-NC3, order Spiral-IV$^X$ is more stable than AABB-AFM order when doping concentration is larger than 0.02 $e$/I, while it is always more preferred than AABB-AFM order for ML-C3 under electron doping.

# APPENDIX J: COMPARISON OF INTERFACIAL INTERACTIONS FOR $NiI_2$/hBN AND $NiI_2$/$SiO_2$ HETEROSTRUCTURES

To more clearly show the different interfacial interactions for ML $NiI_2$ on hBN and $SiO_2$ substrates, we performed calculations for a ML $NiI_2$ on an amorphous $SiO_2$ substrate. The fully relaxed atomic structure of the $NiI_2$/$SiO_2$ interface is shown in Fig. 4a. We plotted line profiles along the $z$ direction of interfacial differential charge density (DCD) variations for both $NiI_2$/hBN (red line in Fig. 4b) and $NiI_2$/$SiO_2$ (blue line in Fig. 4b) interfaces. The $NiI_2$/BN interface exhibits significant charge variations across the interface while that of the $NiI_2$/$SiO_2$ interface is nearly inappreciable, indicating a much weaker interaction of $SiO_2$ to the $NiI_2$ monolayer. The weaker interacting $SiO_2$ also leads to a less-uniformed interfacial structure. As mapped in Fig. 4c, the $SiO_2$ substrate exhibits a much larger range, namely from 2.88 to 5.24 Å, for the vertical distance between interfacial I and O atoms. However, the range for the hBN substrate is much narrower, namely from 3.63 to 3.70 Å, as depicted in Fig. 4d. The smaller vertical corrugation of the $NiI_2$/hBN interface also gives rise to an averaged interfacial distance of 3.67 Å, 0.39 Å smaller than that of the $NiI_2$/$SiO_2$ interface (4.06 Å). Therefore, these results indicate that the amorphous $SiO_2$ substrate, most likely, interacts with the $NiI_2$ ML more weakly than the hBN substrate, which means the $SiO_2$ substrate is less capable of efficiently applying in-plane strain confinements to the $NiI_2$ ML. In other words, the hBN substrate may have a chance to force the $NiI_2$ ML following the $C_3$ symmetry of the substrate, while the $SiO_2$ substrate, most likely, cannot.

# APPENDIX K: EVOLUTION OF LAYER-DEPENDENT MAGNETIC ANISOTROPIC ENERGY (MAE)

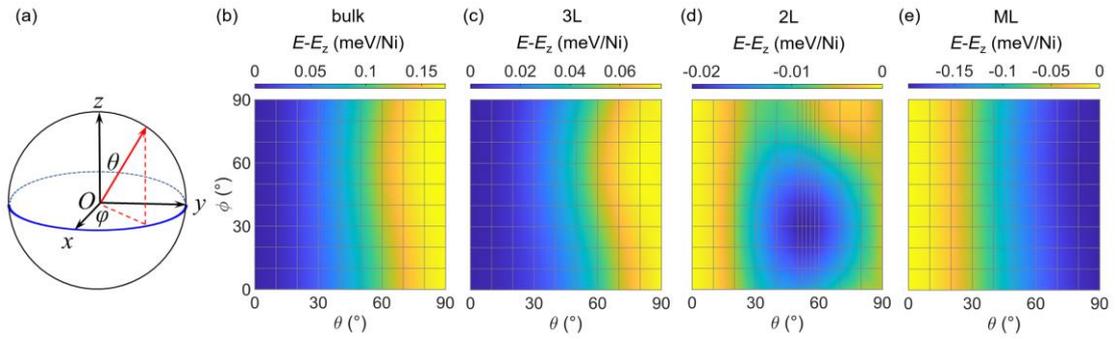

FIG. 21. (a) Illustration of magnetization axes in MAE calculation. Here $\theta$ and $\varphi$ are the angles between the magnetization direction and $z$ and $x$ axes, respectively. Angular dependence energies relative to that of magnetic moment along the $z$ axis for (b) bulk, (c) 3L, (d) 2L and (e) ML $NiI_2$ taking FM order. From ML to bulk $NiI_2$, magnetic moments undergo a spin reorientation from in-plane to out-of-plane direction.

# APPENDIX L: LAYER-DEPENDENT MAGNETIC PARAMETERS

TABLE VII. Isotropic exchange parameters $J_1$, $J_2$, and $J_3$ for the nearest, second nearest, and third nearest Ni atoms, and interlayer isotropic exchange parameters $J_1^\perp$, $J_2^\perp$, and $J_3^\perp$ for the nearest, second nearest, and third nearest Ni atoms, their ratios $|J_1/J_3|$ and $J_2/J_1$, Kitaev $K$, single-ion anisotropy $A_z$, and magnetic dipole interaction over $J_1$ ($B/J_1$), and two-site anisotropy over $J_1$ ($J_{yz}/J_1$) of NiI$_2$ with the C$_3$ symmetry.

|  | $J_1$ (meV) | $J_2$ (meV) | $J_3$ (meV) | $J_1^\perp$ (meV) | $J_2^\perp$ (meV) | $J_3^\perp$ (meV) | $|J_1/J_3|$ | $J_2/J_1$ | $K$ (meV) | $A_z$ (meV) | $B/J_1$ | $J_{yz}/J_1$ |
| --- | --- | --- | --- | --- | --- | --- | --- | --- | --- | --- | --- | --- |
| ML | 3.89 | –0.25 | –3.09 | -- | -- | -- | 1.26 | -0.06 | 0.77 | 1.43 | 0.24 | 0.05 |
| 2L | 3.55 | 0.17 | –3.29 | 0.37 | –1.52 | –0.28 | 1.08 | 0.05 | 2.41 | 1.55 | 0.34 | 0.44 |
| 3L | 3.36 | 0.07 | –3.06 | 0.04 | –1.29 | –0.23 | 1.10 | 0.02 | 2.42 | 1.58 | 0.37 | 0.49 |
| 4L | 3.32 | 0.08 | –3.04 | 0.08 | –1.32 | –0.24 | 1.09 | 0.03 | 2.42 | 1.59 | 0.40 | 0.52 |
| bulk | 3.22 | 0.29 | –3.48 | 0.49 | –2.05 | –0.38 | 0.92 | 0.09 | 2.44 | 1.60 | 0.44 | 0.58 |

TABLE VIII. Isotropic exchange parameters $J_1$, $J_2$, $J_3$, $J_1^\perp$, $J_2^\perp$, and $J_3^\perp$, and their ratios $|J_1/J_3|$ and $J_2/J_1$, $K$, $A_z$, $B/J_1$ and $J_{yz}/J_1$ of NiI$_2$ without the C$_3$ symmetry. The "Strained ML" refers to ML-NC3 undergoing a tensile strain of 1.0 % along the $x$ direction (Strained ML).

| | $J_1$ (meV) | $J_2$ (meV) | $J_3$ (meV) | $J_1^\perp$ (meV) | $J_2^\perp$ (meV) | $J_3^\perp$ (meV) | $|J_1/J_3|$ | $J_2/J_1$ | $K$ (meV) | $A_z$ (meV) | $B/J_1$ | $J_{yz}/J_1$ |
|---|---|---|---|---|---|---|---|---|---|---|---|---|
| ML | 3.92 | −0.26 | −3.20 | -- | -- | -- | 1.22 | -0.07 | 0.77 | 1.43 | 0.24 | 0.05 |
| Strained ML | 3.97 | −0.34 | −3.49 | -- | -- | -- | 1.23 | -0.09 | 1.73 | 1.45 | 0.29 | 0.25 |
| 2L | 3.59 | 0.12 | −3.40 | 0.38 | −1.53 | −0.28 | 1.06 | 0.03 | 2.41 | 1.55 | 0.34 | 0.44 |
| 3L | 3.29 | 0.13 | −3.14 | 0.05 | −1.31 | −0.24 | 1.05 | 0.04 | 2.42 | 1.58 | 0.37 | 0.49 |
| 4L | 3.36 | 0.03 | −3.17 | 0.10 | −1.34 | −0.25 | 1.06 | 0.01 | 2.42 | 1.59 | 0.40 | 0.52 |
| bulk | 3.27 | 0.24 | −3.61 | 0.50 | −2.05 | −0.40 | 0.91 | 0.07 | 2.44 | 1.60 | 0.44 | 0.58 |